# Study of conditions of use of E-services accessible to visually disabled persons

**BOBILLIER CHAUMON M.E**
Laboratoire GRePS
Instit de Psychologie, Université Lyon 2
69676 BRON Cedex (F)
marc-eric.bobillier-chaumon@univ-lyon2.fr

**SANDOZ-GUERMOND F.**
Laboratoire LIESP
INSA de Lyon
69621Villeurbanne Cedex (F)
francoise.sandoz-guermond@insa-lyon.fr

**ABSTRACT**
The aim of this paper is to determine the expectations that French-speaking disabled persons have for electronic administrative sites (utility). At the same time, it is a matter of identifying the difficulties of use that the manipulation of these E-services poses concretely for blind people (usability) and of evaluating the psychosocial impacts on the way of life of these people with specific needs. We show that the lack of numerical accessibility is likely to accentuate the social exclusion of which these people are victim by establishing a numerical glass ceiling.

**ACM Classification Keywords**
Accessibility, Visually disabled persons, E-Government.

**INTRODUCTION & CONTEXT**
The development of new technologies may prove to be a tremendous springboard for the integration of disabled persons (DP) provided that these environments are accessible, usable, and useful; in other words that they take into consideration the various characteristics of the activity and the needs and particularities (cognitive, perceptive, or motive) related to the disability of the users (7, 9).

This question is even more pertinent in the context of quasi-generalized media coverage of the service relationship (E-administration, E-banking, E-commerce, etc.). Various studies worldwide have shown the very weak respect of accessibility criteria despite the numerous standards (section 508 in the USA, the law concerning digital accessibility of administrative services in France, etc.) or labels (*Blindsurfeur* in Belgium, *See it Right* in England, *Accessiweb* in France, etc.) required during the conception of these online services [5]: more than 75% of the assessed sites present level 1 WAI guideline accessibility flaws [9], meaning that accessibility to these sites is impossible for DP [2, 3, 4].

This is becoming a serious problem insomuch as accessibility seems to be one of the social and political levers playing a role in the amelioration of the quality of life of people with disabilities [6, 8, 10]. Indeed, if on the one hand, accessible Internet sites can allow DP greater autonomy by giving them the possibility to complete various activities by themselves; on the other hand, these technologies are also the source of a new type of social stigmatism because of their lack of technological accessibility. The DP must first ask for help to use the system and perform the act.

The objective of our communication is to determine the real contributions of accessible E-services for visually disabled persons as well as evaluate the repercussions of the lack of digital accessibility to these E-services on this population[1]. This is based on the hypothesis that inaccessible technologies will only confirm the inequalities of access to information and services between able-bodied persons and disabled persons, and could even reinforce and intensify them.

In this perspective, we studied the conditions of use of accessible electronic services.

In this perspective, we propose an original approach to study the conditions of use of electronic services accessible to disabled persons. The methodological approach is indeed both:

- Multidimensional: by diagnosing their utility (adaptation to user expectations), usability (ease of use), accessibility (respect of standards and principles), and acceptability (meaning and stakes attributed to the technologies).

- And comparative: since carried out on two user samples (able-bodied and visually impaired) with various levels of E-service experience (novice to expert).

**METHODS**
Our approach draws on three complementary studies:

- The utility of the sites was studied using an online questionnaire on 439 DP with motive, perceptive, and

---

[1] These results are extracted from research on the digital accessibility of electronic administration (ADELA project) financed by the Minister of Research and New Technology (Ministère Délégué à la Recherche et aux Nouvelles Technologies) (Nov. 2004 to Dec. 2005).

cognitive disabilities in order to determine what the E-services bring to the DP and what the DP expect from them.

- The usability and accessibility of the sites was evaluated[2] with user tests based on 3 scenarios (specified below) and two populations: 10 visually disabled participants (VDP) and 10 sighted participants. The participants had comparable sociobiographical characteristics (age, sex, education, etc.), only the mastery of the Internet varied equally in each group (5 novices and 5 experts). For this confrontation, we wanted to know if the problems encountered by the blind were the same as those of the sighted (general problems of usability), or if the problems were amplified by a choice of technology incompatible with their perceptive limits (problems of accessibility). The data collection tools used were simultaneous verbalisation, observations and a satisfaction questionnaire (adapted from the Wammi grid[3]). The indicators measured were the efficiency (time, frequency and nature of errors, omissions, number of selections/strategies to perform a scenario), satisfaction (score out of 5 on the Wammi scale) and efficacy (pass/fail test).

- The acceptability of E-services was analysed using semi-directive interviews of 8 blind participants. The objective was to determine to what extent these services could transform the practices, contacts, and status of the blind. These interviews were recorded and entirely transcribed. A thematic content analysis was performed on this corpus.

**MAIN RESULTS**

**Study of utility of the sites**

Of the 439 DP who answered the online questionnaire, 52% indicated having help with their classic administrative processes. This is due to difficulties in mobility (33.5%) physical accessibility to the building or administrative hours (30.5%), the complexity of forms (23%), or difficult contact with agents (feelings of "being different") (13%). E-administration thus seems like an alternative solution that, incidentally, 52.4% of participants declared to have already used and 32.4% would like to use. These users benefited from them. The role of these E-services as a facilitating tool (finding information, avoiding going out to fill out forms, etc.) is thus confirmed by 90%. The fact that these electronic services allow the DP to avoid requesting someone's help to perform tasks that are often intimate and personal and that they favour the social integration of the DP by providing the same access as an able-bodied person is underlined by, respectively, 90% and 96% of participants.

For the 40% who refuse to use E-services, this position is principally due to technical and ergonomic causes (*lack of reliability and accessibility of environments, data protection, delay of data processing, etc.*) informational causes (*services not complying with the users' needs, unawareness of services offered*) and personal reasons (*preference for classic modes of access, fear of social isolation, entry errors, etc.*). DP support (*sensitisation, education, etc.*) in the acquisition of E-services would certainly help breakdown these barriers at least in part. Finally, even though 46% were opposed to transforming classic services into E-services, and this despite the benefits indicated above, this position should not be seen as a rejection of innovation, but rather as concern and worry, shared by 60% of participants that their specific needs and profiles would not be sufficiently taken into account in the conception of these technologies.

**Evaluation of the usability and accessibility**

Three scenarios were used for these tests: information retrieval from the ANPE (French national employment agency) site (Scenario 1: informational), participation in a public forum (Scenario 2: interactive) and filling out an online form on the Nancy les Vandoeuvre municipal site (Scenario 3: transactional).

|  | Efficacy (% of success in the scenario) | | Satisfaction (mean score / 5) | | Efficiency | | | | | |
|---|---|---|---|---|---|---|---|---|---|---|
|  |  |  |  |  | Mean exploration time (sec) | | Mean number of strategies deployed | | Mean number of selections per scenario | |
|  | Sighted (S) | Blind (B) | S | B | S | B | S | B | S | B |
| Scen. 1 | 100% | 60% | 4.17 | 3.42 | 105 | 814 | 1.38 | 3.40 | 4.38 | 8.20 |
| Scen. 2 | 62.5% | 20% | 2.84 | 2.86 | 230 | 1134 | 2.29 | 3.70 | 6.43 | 7.30 |
| Scen. 3 | 66 % | 10% | 2.84 | 2.86 | 334 | 1176 | 3.00 | 3.44 | 10.83 | 8.22 |

**Table 1: Main results of user tests**

From these analyses (Cf. Table 1), large divergences between the two populations emerge concerning the usability of E-services, as would be expected. The efficacy and efficiency are thus lower for the blind participants than for the sighted participants (with the performances, notably the time, that are up to seven times superior to those of the sighted). However, the satisfaction is globally the same for both groups. We even note a surprising result concerning scenario 3 where the efficiency (for the strategy and selection) is almost advantageous for the blind participants. This piece of data could be explained by a learning effect since the users performed scenarios 2 and 3 on the same site. So, it is the blind expert participants who exploited this learning the best, undoubtedly being used to taking advantage of each action to compensate for their disability.

We note moreover that the usage difficulties penalize mainly the blind the least habituated. The novice blind users seem, in fact, extremely resourceless in dealing with the problem of accessibility of the interface whereas the expert blind users, from their practice and their experience, solicit mental models to compensate for the ergonomic deficiencies of the tool. We can therefore observe a recourse to such schemas when certain blind users anticipate the display of information or interpret inexplicit or polysemous wording by calling on their navigation habits: "*Normally, we should find this information by clicking here…*" On the level of navigation strategies, we can observe that novices opted more often to use search engines to enter the key words of the scenarios to perform (on average 4 of 5 novices) whereas the experts preferred going to the home page to systematically read the proposed links with a voice synthesiser (3 experts of 5). The results show that the failures are more frequent for novices because the key words entered in the search engine are often vague and imprecise. This strategy, which we could qualify as heuristic, is less efficient than the experts' more systematic and general strategy: their mastery of the Jaws system allowed them indeed to consult different the different links very quickly and their experience with E-services also gives them the possibility to promptly locate the most pertinent elements to reach their goal.

---

[2] Ergonomic inspections of accessibility were also performed during the research but won't be presented here due to lack of room.

[3] http://www.wammi.com/using.html

These usage problems come specifically from the choice of conception that does not take into account the perceptive limits of the level of participants, and more generally the principles of accessibility: for example, we can cite newly opening contextual menus remaining unsignalled the appearance of contextual menus not signalled, the density of information presented (over 84 links on a single opening page of a municipal site), the absence of textual alternatives to images, the incoherent structure of pages organised in table format, the use of javascript which makes the screen reader used (Jaws) obsolete, insufficiently explicit links (with do not consider the remaining text content), the opening of new windows not signalled, etc.

Other difficulties common to both groups show, instead, a lack of ergonomics of the sites (according to [1]). It is mainly a matter of certain polysemic terms (*Téléprocédures~Téléservices*), of confusing visited and non-visited links, of the non-deactivation of links on the current page, of unclear error messages, of the dynamic reorganisation of the menus from one page to another, etc. In the end, these results prove that these sites do not take into consideration the inabilities of VDP, and specifically for E-services novices. The accessibility to certain content is very difficult, short of impossible; but moreover, the use of E-services generates a greater mental load that hinders all involvement in the process (shown by the mediocre level of efficiency and by the efforts made to overcome the obstacles to use).

**Analysis of acceptability**

The thematic analysis performed on these data brought out several themes grouped into contributions and risks related to the use of E-services (cf. Table 2 below).

| Impact of E-services on the lives of VDP | References to the theme in 8 interviews | E-services perceived more as a source of improvement | E-services perceived more as a source of deterioration |
|---|---|---|---|
| *Social dimension* | 21 | **13 (62%)** Autonomy, social integration (*through equal access*), Social recognition (*acting without help like able-bodied persons*) | **8 (38%)** Disembodied relation to machines (*absence of personalised attention and consideration*) Risk of social isolation and fear of social exclusion brought on by a digital exclusion |
| *Psychological dimension* | 12 | **7 (58%)** Self-esteem, evaluation (*being able to fend for one-self*) Conservation of confidentiality and privacy of personal information | **5 (42%)** Loss of "know-how" of mobility Fear of losing control of information transmitted (*hacking*), or increased control (*cross referencing of information.*) Feeling of helplessness when confronted with an environment perceived as complex |
| *Cognitive dimension* | 11 | **8 (73%)** Ability to read, classify and collect information in a virtual environment "Demystification" of the administrative process through a simplified access Acquisition of an administrative culture | **3 (27%)** Entropy phenomena: Sorting through the mass of information presented Standardised content of E-services and inadequacies to the needs and profiles of the VDP |
| *Instrumental and operational dimension* | 13 | **8 (61%)** Comfort of life: more mobility Possibility for tenfold increase in action, interaction, and information | **5 (39%)** Insufficient digital accessibility |
| **Total** | 57 | **36 (63%)** | **21 (37%)** |

**Table 2: Main results of thematic analyses of acceptability interviews**

Overall, the visually disabled persons questioned felt that the benefits of the E-services far outweighed the difficulties posed (63% to 37%). E-services thus open "spaces of possibility" that allow DP not only to avoid the cognitive and operative constraints (spatial and temporal) due to their disability, but also to regain a certain autonomy and freedom of action.

These new perspectives contribute to their psychological stability and personal fulfilment (self-esteem). Nevertheless, these people do not idealise these new services either since they are well aware of the stakes linked to the lack of accessibility. Therefore, if administrations are not able to better organize their electronic services, there is a great risk of marginalizing people with specific needs even further. From this point of view, the lack of accessibility represents an additional factor of exclusion and an obstacle to the integration of disabled persons.

Conversely, an exclusive and excessive use of these tools could also turn out to be dangerous since leading to a social isolation (*doing everything remotely from home*) and the loss of a know-how of physical mobility combined with a loss of autonomy of the VDP. "*The problem is that staying home, not moving much, that can create a certain isolation but also a small decline in my mobility comfort level. If I stay at home for a long time and do everything on the Internet, at a certain point, there will be certain things I wouldn't have the courage to do anymore*".

**CONCLUSION AND DISCUSSION**

Our study enables us to show that the conditions of use of E-services depend on three principal factors:

Utility factors in such that the proposed E-services must meet the expectations of the visually disabled persons and bring them a real added value through their use (by augmenting their ability to act, interact, and be informed).

Ergonomic factors (usability and accessibility) in which the specificities of visually disabled persons as well as their level of expertise (with the internet and screen readers) are taken into consideration from the conception.

Psychosocial factors of acceptability where the proposed E-services offer the possibility to truly compensate, assist, and valorise visually disabled persons.

These technologies can, in fact, give value to the individual and confirm/reinforce his place in society by providing him autonomy. The mastery of these ICT could therefore result in the modification of his own perception, the redefinition of his relationship with his entourage and the amelioration of his capacity for social integration. However, these contributions could be limited by the choice of conception. We have, in fact, shown that the quality of ergonomics and the insufficient level of accessibility of the interfaces risk frustrating the user's interaction with the administrative sites and in the end hindering their appropriation and acceptance.

Also, in opting for environments that do not take into account the specific needs and aptitudes of disabled persons, the site creators risk establishing a sort of "technological glass ceiling" that prevents their disabled users from using the E-services offered naturally, whereas able-bodied persons do so with no apparent difficulty. This digital exclusion would be amplified by a social exclusion if the services could only be accessed by the technological channel -- as is planned in the law concerning administrative modernisation.

In sum, the digital chasm resulting from the lack of technological accessibility can be addressed as an additional dimension that adds to the social chasms that disabled persons are subjected to and as a factor contributing to their exclusion and their social isolation.